\documentclass[%
 reprint,
 amsmath,amssymb,
 aps,
]{revtex4-2}

\usepackage{graphicx}
\usepackage{dcolumn}
\usepackage{bm}
\usepackage[utf8]{inputenc}
\usepackage{amsmath}
\usepackage{subcaption}
\usepackage{caption}
\usepackage{comment}

\usepackage{hyperref,bm}
\usepackage[capitalize]{cleveref}
\hypersetup{
	colorlinks,
	citecolor=blue,    filecolor=blue,
	linkcolor=blue,     urlcolor=blue
}

\begin{document}

\preprint{APS/123-QED}

\title{Axiparabola: a new tool for high-intensity optics}

\author{Kosta Oubrerie$^1$}
 \email{kosta.oubrerie@ensta-paris.fr}
\author{Igor A. Andriyash$^1$}
\author{Ronan Lahaye$^1$}
\author{Slava Smartsev$^2$}
\author{Victor Malka$^2$}
\author{Cédric Thaury$^1$}
\affiliation{%
 $^1$Laboratoire d'Optique Appliquée, École Polytechnique, ENSTA Paris, CNRS, Institut Polytechnique de Paris, 91762 Palaiseau, FRANCE\\
 $^2$Department of Physics of Complex Systems, Weizmann Institute of Science,\\ Rehovot 7610001, ISRAEL
}%


\begin{abstract}
    An axiparabola is a reflective aspherical optics that focuses a light beam into an extended focal line. The light intensity and group velocity profiles along the focus are adjustable through the proper design. The on-axis light velocity can be controlled, for instance, by adding spatio-temporal couplings via chromatic optics on the incoming beam. Therefore the energy deposition along the axis can be either subluminal or superluminal as required in various applications. This article first explores how the axiparabola design defines its properties in the geometric optics approximation. Then the obtained description is considered in numerical simulations for two cases of interest for laser-plasma acceleration. We show that the axiparabola can be used either to generate a plasma waveguide to overcome diffraction or for driving a dephasingless wakefield accelerator.
\end{abstract}

\maketitle


\section{Introduction}

Bessel beams are diffractionless light waves and can propagate with subluminal or superluminal velocities~\cite{kuntz2009spatial}.
These properties have found many applications in  material processing \cite{meyer2017single}, optical guiding of microscopic particles~\cite{ahlawat2011long}, optical coherence tomography~\cite{ding2002high} and formation of plasma waveguides~\cite{durfee1993light}. 
Several optics, e. g. axilenses~\cite{davidson1991holographic}, axicon lenses~\cite{arlt2000generation}, or conic mirrors, are able to generate quasi-Bessel beams.  The axiparabola is an aspherical mirror, which combines the advantages of these different optics by being achromatic, having a high damage threshold and allowing for control of the intensity distribution along the focal line~\cite{smartsev2019axiparabola}. These features make it the perfect tool for producing ultra-short quasi-Bessel beams at very high intensity.

A beam focused by an axiparabola has phase and group velocities, which are equal to each other, and in vacuum they are larger than the speed of light in vacuum $c$. Depending on the  mirror design, the beam's on-axis velocity can either only grow or only fall along the focal line. Moreover, the intensity and velocity profiles are fully coupled: any change of the optics surface through its so-called \textit{sag function}, aiming at modifying the intensity profile, impacts the velocity profile. However, this restriction changes if the focused beam initially contains spatio-temporal couplings (STC), which allow to decouple the intensity and velocity profiles.

In this article, we study theoretically the properties of a laser beam focused by an axiparabola and we present ways to control these properties by the means of STC. 
We first derive basic equations and describe the intensity and velocity profiles without STC and in the geometrical optics approximation. Secondly, we explore effects of STC on a velocity profile and determine the way to control it. We then present an optical propagation algorithm, which we further use to confirm the predictions of geometric optics. Finally we discuss in more detail two examples of axiparabolas of relevance for laser plasma accelerators development, such as dephasingless wakefield acceleration~\cite{caizergues2020phase,palastro2020dephasingless,debus2019circumventing} or diffractionless wakefield acceleration with an all-optical plasma waveguide~\cite{smartsev2019axiparabola,shalloo2018hydrodynamic,shalloo2019low}. 

\section{Basic equations}

An axiparabola is an aspheric mirror that reflects a collimated beam into an extended focal line by  focusing rays at different focal planes depending on their radial coordinate $r$ on the mirror. The shape of its surface is defined by the sag function $s(r)$. The rays coming parallel to the optical axis $\bm{\zeta}$ at the radial coordinate $r$  impinge the mirror at $\zeta=s(r)$ and are focused at $\zeta=f(r)=f_0+z(r)$ with $f_0$ the nominal focal length, $z(r)\in [0,\delta]$ the focal line coordinate along the $\bm{\zeta}$ axis, and $|\delta |$ the focal depth. The main differences between an axiparabola and an axicon lens, or a conic mirror, is that  $f_0$ is non-zero, and that $\delta$ can be either positive  or negative ($\delta <0$ corresponds to outer rays focused first). This means that the focal spot size and effective Rayleigh range are decoupled. The mean focal spot transverse size is controlled by $f_0$ and the focal range ("effective Rayleigh range") by $\delta$.

\begin{figure}[htbp]
    \centering
    \includegraphics[width=0.35\textwidth]{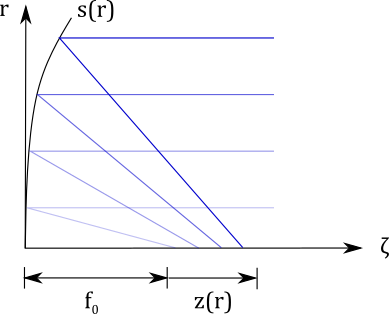}
    \label{fig:axiparabola_rays}
    \caption{Schematic representation of rays focused by an axiparabola with a positive z(r).}
\end{figure}

From geometrical optics laws, the sag function is defined by~\cite{smartsev2019axiparabola}
\begin{equation}
s(r) + \frac{r}{2}\left(\frac{1}{D(r)}-D(r)\right) = f(r)
\end{equation}
with $D(r)=ds/dr$ the sag derivative with respect to the radial coordinate. Computing $D(r)$ and choosing the positive solution, we get 
\begin{equation}
 r \frac{ds}{dr} =s(r ) - f (r )+ \sqrt{[s(r ) -f (r )]^2 + r^2}\,.
 \label{diff_eq_s}
\end{equation}
Let $\sigma(r)=s(r)- r^2/4f_0$ be the deviation to a perfect parabola and $R$ the beam radius. 
Assuming that the deviation to a parabola is small, $\sigma \ll r^2/4f_0$ and $\delta^2/8R^2 \ll 1$, Eq.~\eqref{diff_eq_s} can be approximated as
\begin{equation}
    \frac{d\sigma(r)}{dr}=-\frac{rz(r)}{2ff_0}+o(r^3)\,.
    \label{dsigm}
\end{equation}
The integration of Eqs~\eqref{diff_eq_s} or~\eqref{dsigm} allows to determine the required sag function for achieving a given focal line $f(r)$. This will fix the intensity distribution at focus as well as the light velocity evolution along the focal line. 

\subsection{Longitudinal intensity distribution}

A key parameter for the practical use of axiparabolas is the intensity distribution along the propagation axis. Let us start from the geometrical optics description and define $\lambda_z(z)$ as the linear density of ray along the propagation axis $\bm{\zeta}$ (in $\text{W.m}^{-1}$). The laser intensity on the axiparabola $I_0(r)$ (in $\text{W.m}^{-2}$) is defined as the surface density of rays, and it is related to the linear ray density as
\begin{equation}
\lambda_z(z) dz =2 \pi r dr I_0(r)\label{inten} \,.
\end{equation}
In the present study we assume a top-hat beam profile, so that the intensity on the mirror is uniform, $I_0(r)=P_0/(\pi R^2)$ with  $P_0$ the incident power, leading to
\begin{equation}
\frac{d r}{d z} =\frac{\lambda_z(z)R^2}{2 P_0 r}\mathtt{,}\label{drdz}
\end{equation}
and finally
\begin{equation}
r(z)={R}\left(\int_0^z\frac{\lambda_z(z')}{P_0}dz'\right)^{1/2}\,.\label{r_z}
\end{equation}
This equation allows to calculate the function $r(z)$ and hence the sag function defining the desired intensity profile $\lambda_z(z)$. For instance, for a focal line of constant intensity, $\lambda_z=P_0/\delta$,  we get that    $f=f_0+\delta r^2/R^2$. The sag function can then  be obtained by integration of Eq.\eqref{diff_eq_s}.

\subsection{Transverse intensity distribution}
While for a classical focusing optic, the focal depth, or in other words the Rayleigh length, is closely linked to the beam waist, these two quantities are decoupled at the focus of an axiparabola. 
For the sake of simplicity, we illustrate  this property by considering a top-hat incident beam.
In the Fresnel diffraction regime, the field at the distance $f(z)=f_0+z$ from the axiparabola is

\begin{equation}
\begin{split}
E(r_{\zeta},z)=&-i\cfrac{ E_0 k}{f(z)}\;e^{i k\left(f(z)+\frac{r_{\zeta}^2}{2f(z)}\right)}\;\\&\times\int_0^R dr e^{i \Psi(r)} r J_0\left( k \cfrac{r_{\zeta} r}{f(z)}\right)
\end{split}
\end{equation}
with $k=2\pi/\lambda$ the wave-vector, $r_{\zeta}$ the radial coordinate over the focal line, $J_0$ the first Bessel function of first kind and 
\begin{equation}
\begin{split}
    \Psi(x)&=k\left( x^2 /2f(x)-2s(x)\right)\\&= k\left(x^2 /2f(x)-x^2/2f_0-2\sigma(x)\right)\\&\approx -k\left(x^2 d/2f_0^2+2\sigma(x)\right)
\end{split}
\end{equation}
for $z\ll f_0$. For $r\gg 1/k$ we can use the stationary phase method to estimate the integral:
\begin{equation}
\begin{split}
E(r_{\zeta},z)=&-i\frac{ E_0 k}{f(z)}e^{i k\left(f(z)+\frac{r_{\zeta}^2}{2f(z)}\right)} \sqrt{\frac{2\pi}{\Psi''(r_s)}}\\&\times r_s J_0\left( k \frac{r_{\zeta} r_s}{f(z)}\right)e^{i \Psi(r_s)+i \pi/4}
\end{split}
\end{equation}
with $r_s$ the coordinate such as $\Psi'(r_s)=0$ (note that we assumed $\Psi''(r_s)>0$). According to Eq.~\eqref{dsigm}, we have 
\begin{align}
\Psi'(x)\approx-k\frac{x z}{f_0^2}+k \frac{x z(x)}{f_0^2}+ o\left(k x^3/f_0^2\right)
\end{align}
It follows from Eq.~\eqref{drdz} that $\Psi''(r_s)= 2 r_s^2 P_0/ \left(\lambda_z(z) R^2 f_0^2\right)$.
As a consequence, the intensity along the focal line is
 \begin{equation}
 \begin{split}
I(r_{\zeta},z)&=|E(r_{\zeta},z)|^2\\
&=\frac{ E_0^2 k^2}{f_0^2} \frac{2\pi}{k\Psi''(r_s)}r_s^2 J_0^2\left( k \frac{r_{\zeta} r_s}{f_0}\right)\\
&=k \lambda_z(z)J_0^2\left( k \frac{r_{\zeta} r_s}{f_0}\right)\mathtt{,} \label{eq:I_rz}
\end{split}
\end{equation}
with
\begin{align}
r_s=R\sqrt{\int_0^z{\frac{ \lambda_z(z')}{P_0}dz'}}
\end{align}
One may easily see that the radial intensity profile is described by the first Bessel function and that the on-axis intensity is $I_0(z)=I(0,z)=k\lambda_z(z)$. We finally find that the first-zero radius, for $r_s \gg 1/k$, is 
\begin{align}
r_{\zeta,0}(z)\approx 0.77 \lambda N \left(\int_0^z \frac{I_0(z')}{k P_0}dz' \right)^{-1/2}\mathtt{,}\label{eq:r_axi}
\end{align}
with $N=f_0/2R$ the f-number. As a result, Eqs.~\eqref{eq:I_rz} and~\eqref{eq:r_axi} show that the intensity does not depend on $N$, and hence that $r_{0}$ can be adjusted independently of  $I_0$ by changing $N$. For example, for a constant intensity focal line we get $I_0=kP_0/\delta$ and $r_{\zeta,0}=0.77\lambda N (\delta/z)^{1/2}$; the intensity at focus depends only on the beam power and focal depth, while the focal spot  is a function of $N$. Therefore an axiparabola can redistribute the laser energy into a focal line combining a long focal depth and a very small focal spot.

\subsection{Velocity profile}
It is well-known that Bessel beams travel at constant velocities that can exceed light speed in vacuum. However, as shown in previous section, an axiparabola generates a quasi-Bessel beam, for which the longitudinal group velocity is still superluminal but is no longer constant. Defining the group velocity of the beam as the velocity of the intensity peak along the focal line we can describe it using Eq.~\eqref{diff_eq_s}.
The optical path of light in vacuum from the axiparabola to the optical axis is 
\begin{equation}
 p(r) = \sqrt{[s(r ) -f (r )]^2 + r^2} - s(r)\,.
 \label{optical_path}
\end{equation}
The geometrical group velocity is the change of the focus position in time $v = df/dt$, and we note that the increase of the optical path on axis $\zeta$ is $dp = c dt$. With that in mind, we can parametrize differentials as functions of $r$, and express the group velocity as:
\begin{equation}
\frac{v}{c} = \frac{df}{dr}\,\left(\frac{dp}{dr}\right)^{-1}=\left(\frac{dp}{dz}\right)^{-1}\,.
\label{v1}
\end{equation}
Then using Eqs.~\eqref{diff_eq_s} and~\eqref{optical_path}, we get in the paraxial limit
\begin{equation}
\frac{v}{c} = 1 + \frac{2\left(\frac{ds}{dr}\right)^2}{1-\left(\frac{ds}{dr}\right)^2} = 1 + \frac{r^2}{2f^2}\,.
\label{velocity}
\end{equation}
Eq.~\eqref{velocity} shows that the group velocity is always larger than the speed of light in vacuum and that its  evolution along the focal line can be either increasing or decreasing, depending on whether $\delta$ is  positive or negative respectively. For a top-hat incident beam, we get from Eq.~\eqref{r_z} that the group velocity of the focal line is
\begin{equation}
\frac{v}{c} = 1 + \frac{R^2}{2f^2 P_0}\int_0^z\lambda_z(z')d z'\,,
\label{velocity2}
\end{equation}
which illustrates the direct relation of the group velocity to the local intensity. This link between velocity and intensity hinders the actual ability of axiparabolas to control the velocity of laser power propagation. Nevertheless, as will be shown later, this scheme still holds the opportunity to dissociate that connection through spatio-temporal couplings, which gives another degree of freedom to control and modify the group velocity along the focal line.

\begin{figure}[t]
    \centering
    \includegraphics[width=0.35\textwidth]{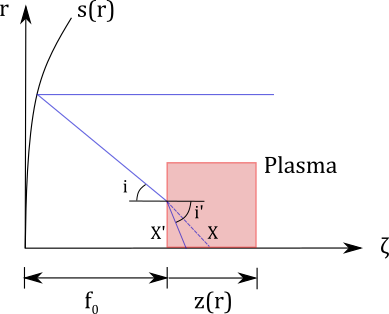}
    \caption{Schematic representation of a ray focused by an axiparabola with a positive z(r) in a constant density plasma.}
    \label{fig:axiparabola_interface}
\end{figure}

\section{Control of the velocity}

\subsection{Group velocity}

Axiparabola focuses different annular beamlets annuli to different focal planes depending on their incident radial coordinates. This spatial separation allows for control of the beamlets arrival and thus control of the group velocity. In other words: group velocity along the focal line depends on the radial coordinate on the mirror. Hence, its value can be modified  by adding a radial delay prior to the axiparabola.
This can be exemplified by considering a linear propagation in vacuum. 
Starting from Eq.~\eqref{v1}, a radial delay $\tau(r)$ is added and modifies the velocity $v$ as $v_{m}$ :
\begin{equation}
  \frac{v_{m}}{c} = \left(\frac{d\left(p+c\tau\right)}{dz'}\right)^{-1} \,.
\end{equation}
Assuming that $p \gg c\tau$, this leads to 
\begin{equation}
    \frac{v_{m}}{c} \simeq \frac{v}{c}\left(1-\frac{v}{c}c\frac{d\tau}{dr}\frac{dr}{dz}\right)\,.
    \label{retardation_2}
\end{equation}
This highlights that the velocity can be controlled by introducing simple spatio-temporal couplings. 
To properly illustrate this phenomenon, let's assume a top-hat beam in the paraxial limit for an axiparabola such as $r^2/2f^2 \ll 1$: the radial delay needed for having an intensity peak that propagates, at a constant velocity $c+v_0$ with $v_0\ll c$, is
\begin{equation}
    c\frac{d\tau}{dr}\simeq\left(\frac{v}{c}-\frac{v_0}{c}-1\right)\left(\frac{c}{v}\right)^2\frac{2P_0r}{\lambda_zR^2} \,,
    \label{retardation_1}
\end{equation}
leading to 
\begin{align}
    c\tau\simeq\frac{P_0}{\lambda_zR^2}\left( -\frac{v_0}{c} r^2 + \frac{1}{2f^2}\left( \frac{v_0}{c} + \frac{1}{2} \right) r^4\right) + o(r^5) \,.
\end{align}
Here the term $\propto r^4$  flattens the velocity profile to get an intensity peak that propagates at a constant velocity $c$, while the quadratic term allows to adjust the value of the velocity around $c$. This quadratic term corresponds to the pulse front curvature (also known as longitudinal chromatic aberration), an aberration which is present in many laser chains, and which can be controlled by using simple plano-convex optics in the laser chain~\cite{Cui:19,kabacinski2021measurement}. Achieving the  $r^4$ term would require the use of aspheric lenses, specially designed for a given axiparabola. 

This simple prediction model can also be adjusted to take into account the medium in which the laser propagation occurs depending on the applications. For applications in the field of laser-plasma acceleration, the design has also to account for the laser propagation in plasma. For this, let us consider a uniform plasma slab localized in between the focal line's boundaries, i.e. plasma density $n_e$ is constant for $f_0 \leq z \leq f_0+\delta$ and zero elsewhere, as shown in Fig.~\ref{fig:axiparabola_interface}. The plasma is assumed to be underdense, which means $n_e\ll n_c$, with $n_c=\pi/\left(\lambda_0^2 r_e\right) = 1.1\cdot 10^{21}\, (\lambda_0 [\mu\text{m}])^{-2}$~cm$^{-3}$ being the critical plasma density for the wavelength $\lambda_0$ and $r_e$ the classical electron radius. 

Propagation of light in plasma is affected by the refraction at the vacuum/plasma interface and by the modification of the light velocity in plasma. Let $i(r)$ be the angle between the optical axis and the rays that are focused at $z(r)$ in vacuum and $X(r)$ the propagation distance after the vacuum/plasma interface of the rays that are focused at $z(r)$ in vacuum. Assuming paraxial rays, these two variables are defined by
\begin{equation}
    i(r)=\arctan(r/(f-s))
\end{equation}
\begin{equation}
    z=X\cos(i) \,.
\end{equation}
The rays that are focused at $z(r)$ in vacuum cross the optical axis at a new coordinate
\begin{equation}
    \begin{split}
        z' & = X'\cos(i')\\
        & \simeq z\eta\left(1+\frac{i^2}{2}\right)
    \end{split}
\end{equation}
for $i\ll 1$ and with $\eta \simeq 1-n_e/(2n_c)$ being the refractive index of plasma. This involves a shortening of the focal line $\left(\delta'<\delta\right)$ that leads to a corresponding increase of the intensity. As the optical path in plasma remains equal to the one in vacuum $\left(X = X'/\eta\right)$, the decrease of the propagation distance is compensated by the slower group velocity of light in plasma $\left(v_g/c = 1 - n_e/(2n_c)\right)$. The group velocity of the focal line in plasma can therefore be written 
\begin{equation}
    v_p=v\frac{dz'}{dz} \,.
\end{equation}
with $v_p$ the group velocity in plasma and $v$ the one in vacuum.
Following the same method as in Eq.~\eqref{retardation_2}, this involves that the modified velocity in plasma can be written
\begin{equation}
    \frac{v_{p,m}}{c} \simeq \frac{v_p}{c}\left(1-\frac{v}{c}c\frac{d\tau}{dr}\frac{dr}{dz}\right)\,.
\end{equation}
From this equation, the required radial delay can be computed with the same process as in Eq.~\eqref{retardation_1}.
This study of the group velocity is applicable in vacuum and in any transparent medium and shows that the group velocity can be adjusted independently of the intensity  of the focal line,  allowing for subluminal or superluminal velocities.

\subsection{Phase velocity}

In dispersive media, group velocity and phase velocity can be different. Therefore, to fully describe the focal line propagation, the impact of the control of the group velocity through STC on the phase velocity is also of interest.  Let $\varphi(z,t) = k p(z)-\omega t$  be the beam phase, with $\omega$ the laser pulsation.
The phase velocity is 
\begin{equation}
    v_{\varphi} = \frac{d\varphi/dt}{d\varphi/dz} \,,
\end{equation}
and in a plasma and in the absence of STC, its  spatial derivative can be written
\begin{equation}
        \frac{d\varphi}{dz}  = k \frac{dp}{dz} = \frac{\omega}{v}
\end{equation}
with $v$ the group velocity.
This leads to the following formula :
\begin{equation}
    v_{\varphi} = v = 1 +\frac{r^2}{2f^2} \,.
\end{equation}
The phase velocity of the focal line is thus equal to its group velocity. Now let us observe the evolution of the phase velocity when a radial delay $\tau(r)$ is added prior to the axiparabola as presented in the previous subsection. The phase is then changed to 
\begin{equation}
    \begin{split}
        \varphi_m & = k\left( p+c\tau \right) - \omega \left( t + \tau \right)\\
        & = kp-\omega t = \varphi \,.
    \end{split}
\end{equation}
This means that the focal line phase is not modified by the introduction of a radial delay and therefore the phase velocity is always equal to the unaltered group velocity, and thus different from the group velocity in presence of STC:
\begin{equation}
    v_{\varphi} = v = 1 +\frac{r^2}{2f^2} \,.
\end{equation}

\section{Optical propagation modeling}

To simulate the evolution of the laser field along its path we solve numerically the Helmholtz equation. In the Fourier space, propagation of the complex field $\psi (\omega, k_x, k_y, z)$ from the plane $z_0$ to $z_1$ can be computed by multiplying it by the propagator 
$${\psi_1 = \psi_0\, \exp\left( i (z_1-z_0) \sqrt{\omega^2/c^2 - k_x^2 - k_y^2} \right)}\,.$$
Here, the field is considered strictly cylindrically symmetric, and solutions can be expressed via cylindric modes, i.e. Bessel functions $\psi(r) = \int r\mathrm{d}r \hat{\psi} J_0(k_r r)$, where  $k_r$ is equivalent to $\sqrt{k_x^2 + k_y^2}$ in the propagator expression.

One method, based on the quasi-discrete Hankel transform (QDHT) was demonstrated in \cite{guizar:JOCA2004}. The approach was based on the symmetric transform (same matrix for forward and inverse projections), where both spatial and spectral  axes, $r$ and $k_r$, were built on the zeros of $J_0$. In the case of a sharply focused beam, the beam waist can be $10^2-10^3$ times smaller than the spot on the mirror, and to resolve both one may require large numbers of points along the radial and spectral axes $N_r = N_{k_r} \gtrsim 10^4$. 

For our calculations, we have used a non-symmetric transform with different sampling of the initial and focused images. For this we consider field decomposition into the series, $\psi(r_i) = \sum_{j=0}^{N_r-1}   \hat{\psi}_j J_0(k_{r,j} r_i)$, where $r_i = R_\mathrm{max} \alpha_i/\alpha_{N_r}$ and $k_{r,j} = \alpha_j/ R_\mathrm{max}$ with $\alpha_i$ defined as the roots of Bessel function $J_0$ (see \cite{guizar:JOCA2004}). This gives the inverse Hankel transform matrix $T^{(-1)}_{ji} = J_0(\alpha_i \alpha_j/ \alpha_{N_r} )$, and the forward transform $T_{ij}$,  which is found by the numerical inversion of $T^{(-1)}_{ji}$. To reconstruct the field, we use the re-sampled inverse transform  $\overline{T^{(-1)}}_{ji} = J(r'_i k_{r,j} )$, where axis $k_{r,j}$ is same as in $T_{ij}$, but $r'_i$ is sampled uniformly in a small area around the beam effective waist. 

Both schemes have been numerically in all relevant cases. The resampling scheme demonstrated a very good agreement with the original approach \cite{guizar:JOCA2004} with significant sampling reduction (reduction $\sim 8$~times of $N_r$). The implementation of this and a few other schemes can be found in the open-source library "\texttt{Axiprop}" in \cite{axiprop:2020}.

\section{Axiparabola with a constant intensity focal line}
\label{sec:cons_i}

Let us now consider an axiparabola design for relevant applications in laser-plasma acceleration. In laser wakefield accelerators (LPA), an ultra-short laser pulse is focused in a plasma to generate a plasma wave. The amplitude of the longitudinal electric field of such a wave can be a few orders of magnitude higher than those created in conventional linear accelerators. One fundamental limitation of LPA is the particle-wave dephasing that is due to the mismatch between the group velocity of the laser in plasma and the velocity of relativistic electrons. An axiparabola with a constant intensity line could be used to accelerate electrons and overcome this limit. As shown in Eq.~\eqref{retardation_1}, the combination of axiparabola and appropriate spatio-temporal couplings allows to control the group velocity, and hence eventually to phase-lock the light beam velocity on the electron beam velocity. This paved the way for a new acceleration concept that could increase the energy of the generated electrons by at least an order of magnitude~\cite{caizergues2020phase,palastro2020dephasingless}.


To design an axiparabola with a constant intensity focal line, it was assumed that the linear density of rays $\lambda_z=P_0/\delta$, with $\delta$ the focal line length. By replacing the expression of $\lambda_z$ in Eq.~\eqref{r_z}, the focal length expression becomes 
\begin{align}
    f=f_0+\delta \left(\frac{r}{R}\right)^2 \,.
    \label{focal_int}
\end{align}

\begin{figure}[htbp]
    \centering
    \includegraphics[width=0.45\textwidth]{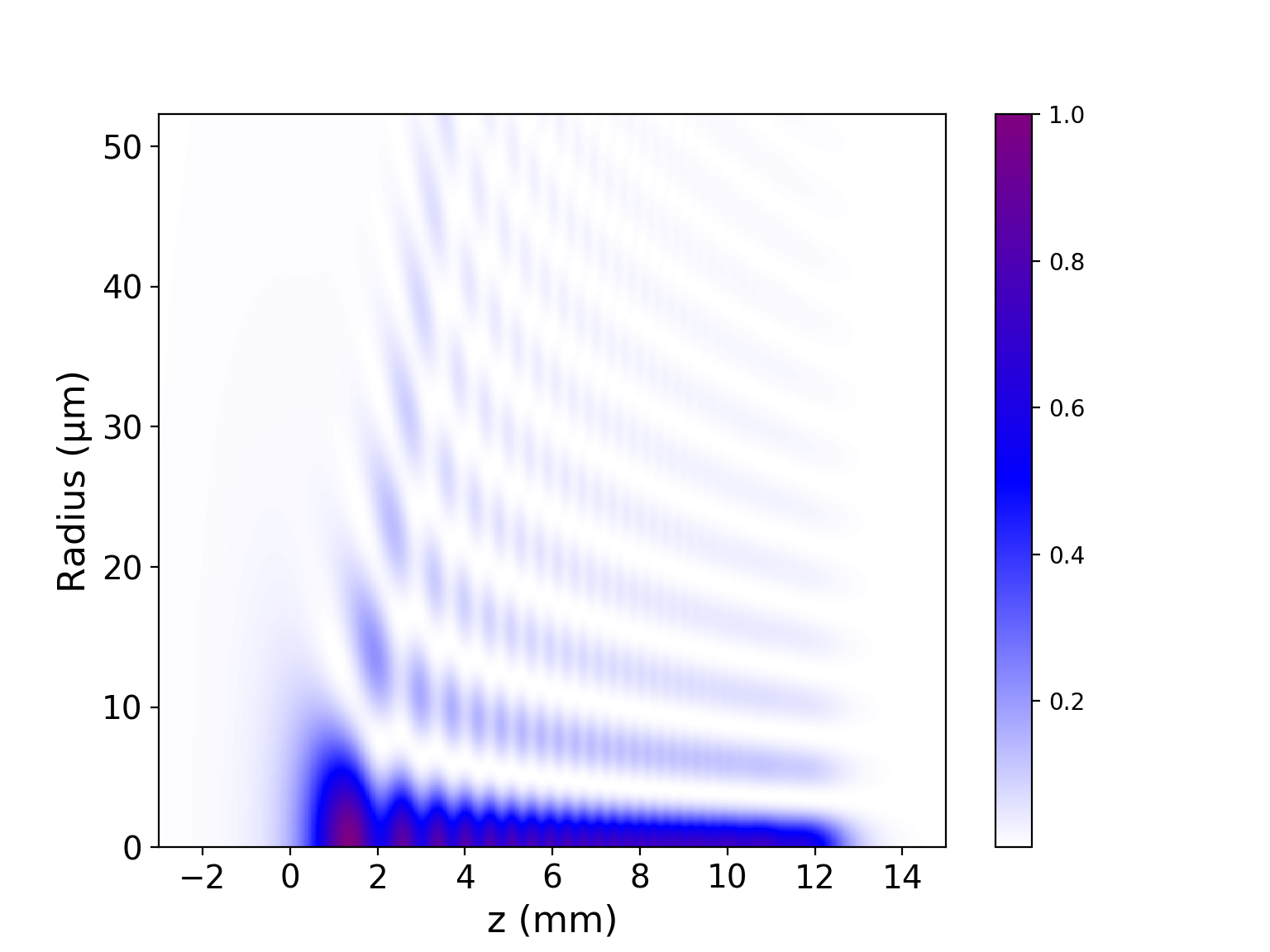}
    \caption{Intensity map of a constant intensity focal line in arbitrary units, as a function of $(r,z)$.}
    \label{fig:foc_int}
\end{figure}
\begin{figure}[t]
    \centering
    \includegraphics[width=0.45\textwidth]{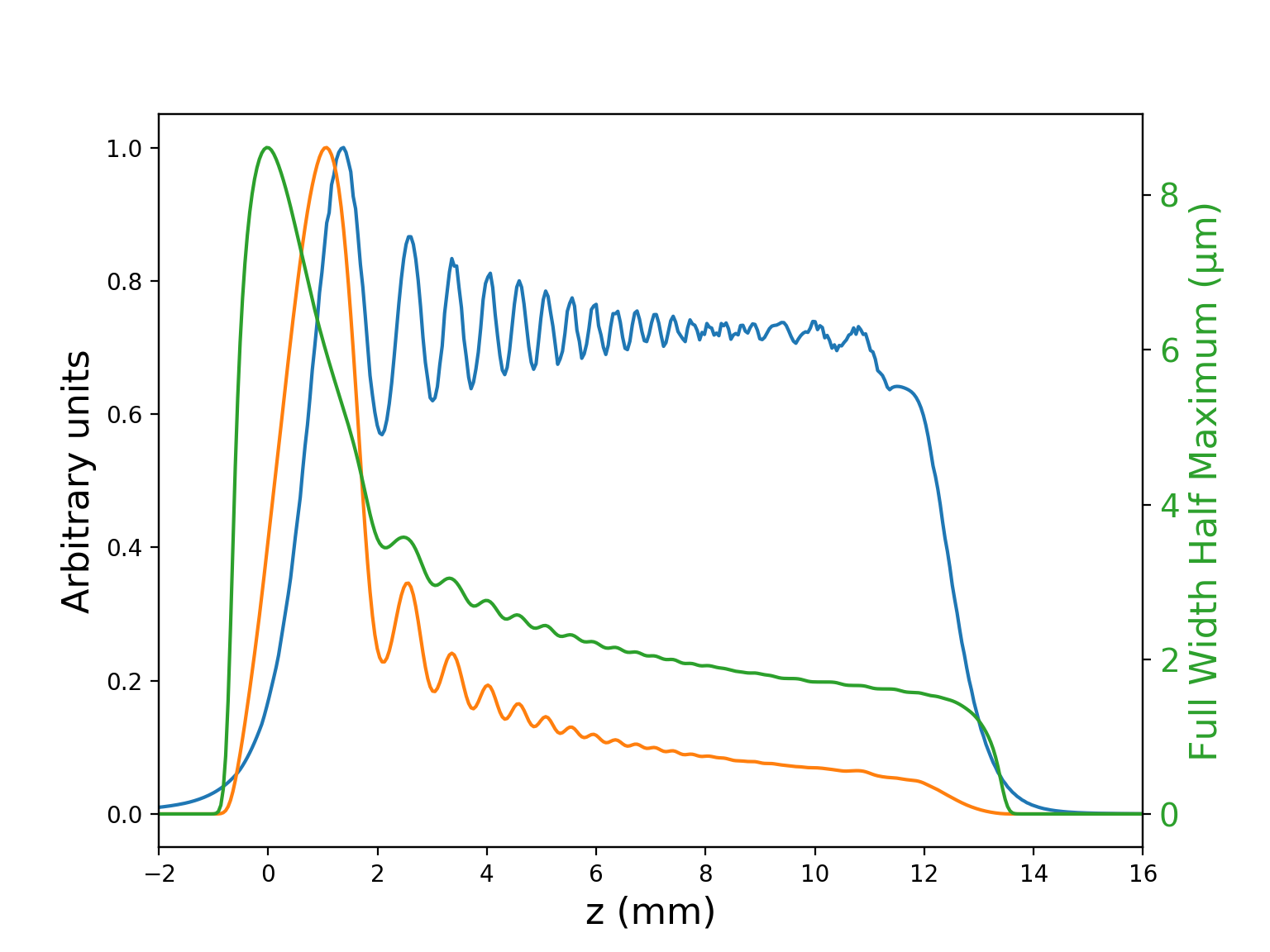}
    \caption{Axiparabola relative intensity (blue curve), relative energy (orange curve) and first-zero radius (green curve), along the focal line.}
    \label{fig:int_int}
\end{figure}

From Eq.~\eqref{velocity2}, the group velocity can now be written 
\begin{align}
\frac{v}{c}=1 +\frac{R^2}{2\delta f^2}z\,.
\label{velocity_int}
\end{align}
Note that the group velocity with this particular axiparabola design has a linear dependence on the  position along focal line $z$. For simulations, the following characteristics were chosen: a nominal focus $f_0=400$ mm, a focal line length $\delta=15$ mm and a radius $R=38.1$ mm.

In Fig.~\ref{fig:foc_int}, we plot the radial distribution of laser field intensity mapped along its propagation, and Fig.~\ref{fig:int_int} shows the beam characteristics. From Fig.~\ref{fig:int_int} one can see that, in agreement with theoretical considerations laser intensity remains constant along the focal line.  The sinusoidal variations are typical characteristics of a Bessel beam. As the first-zero radius diminishes along the focal line, while the intensity remains constant, the energy encircled in the focal spot also diminishes proportionally to the first-zero radius.

\begin{figure}[h]
    \centering
    \includegraphics[width=0.45\textwidth]{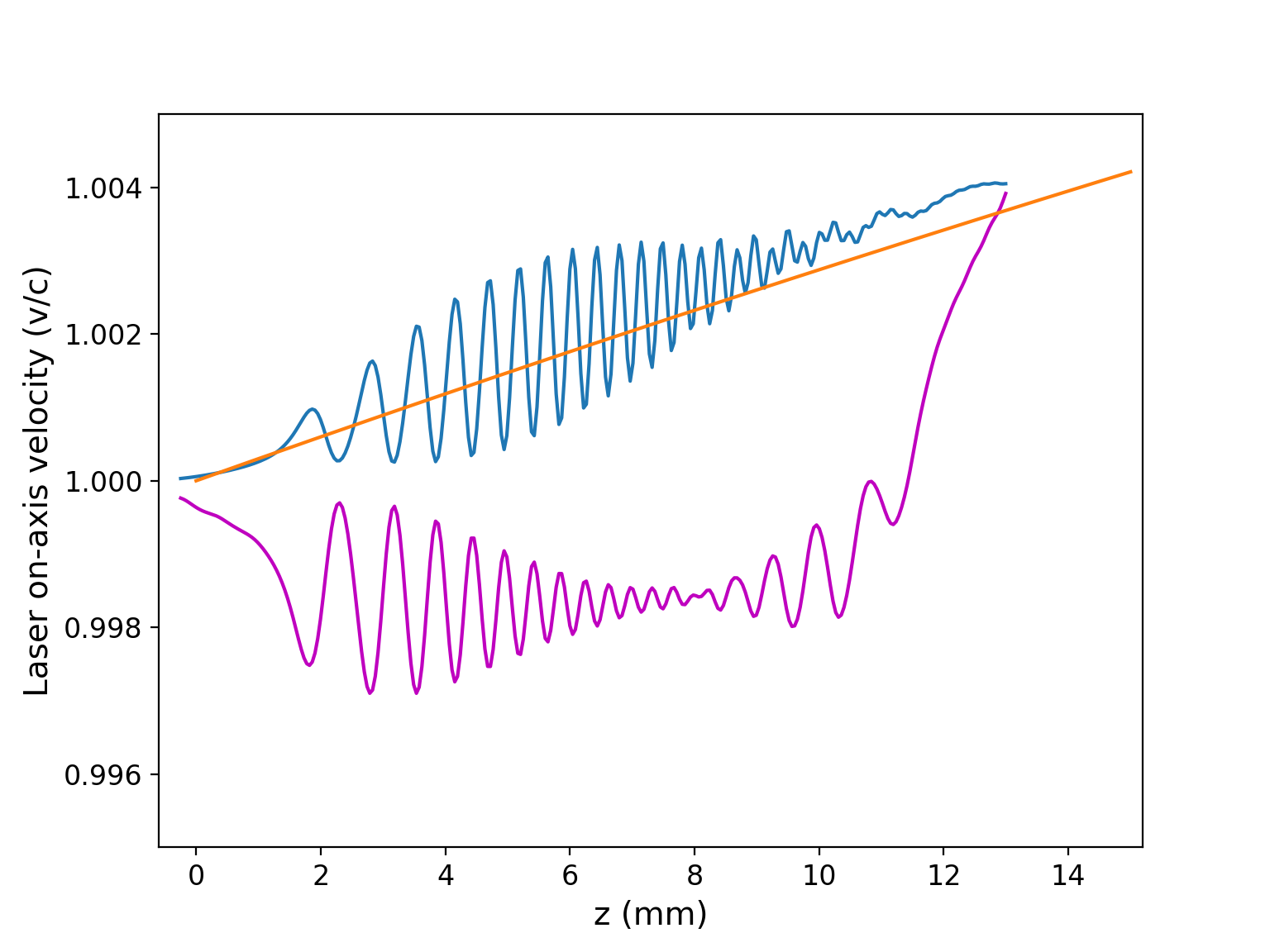}
    \caption{Group velocity as a function of the position along the focal line of the axiparabola. The orange curve corresponds to Eq.~\eqref{velocity_int}, the blue and purple ones to simulation data obtained without and with the radial delay displayed in Fig~\ref{fig:int_ret}, respectively.}
    \label{fig:vel_int}
\end{figure}

\begin{figure}[t]
    \centering
    \includegraphics[height=6cm]{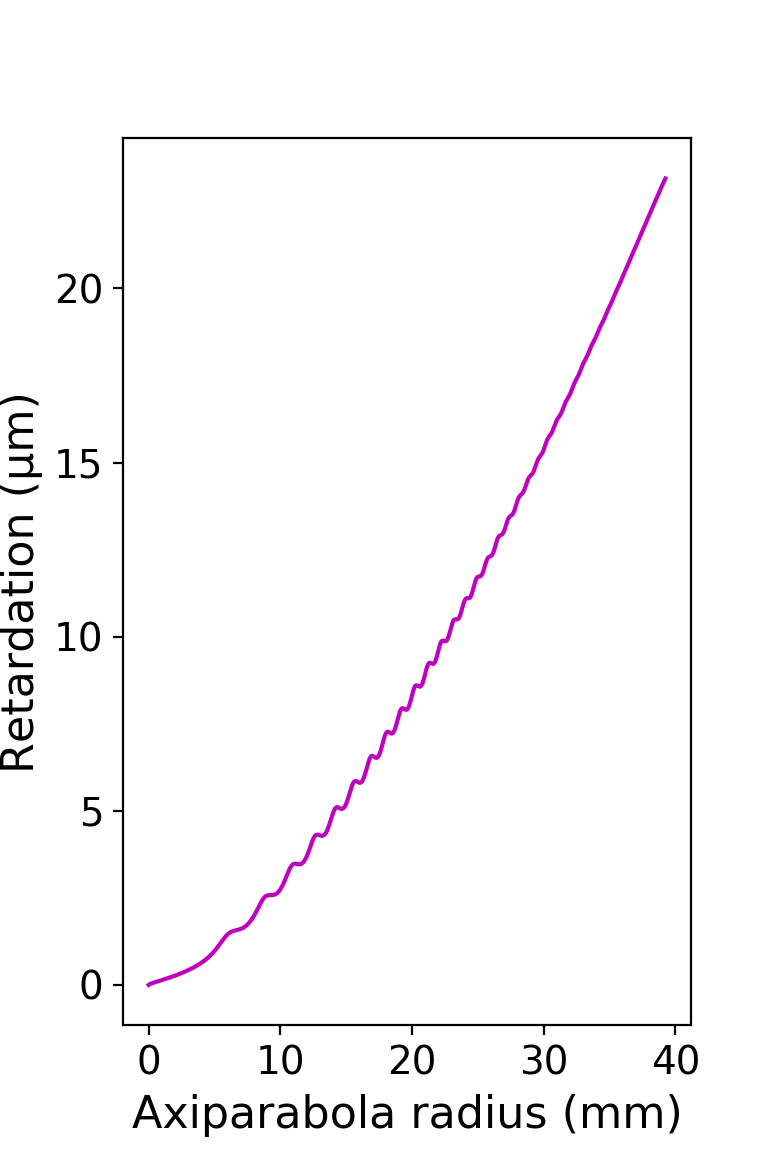}
    \caption{Radial delay needed for a group velocity equal to c, from Eq.~\eqref{retardation_1}}
    \label{fig:int_ret}
\end{figure}

The group velocity of the focal line is calculated by averaging over the intensity map, hence oscillations are visible within the group velocity's evolution along the focal line (Fig.~\ref{fig:vel_int}). Leaving aside the oscillations, which can not be described in the framework of geometrical optics,  the group velocity increases linearly with $z$, as expected from Eq.~\eqref{velocity_int}. The spatio-temporal couplings computed from Eq.~\eqref{retardation_1} and shown in Fig.~\ref{fig:int_ret}, enable to obtain a focal line with a constant group velocity close to the light velocity in vacuum c. The gap observed between the obtained and aimed group velocities, as well as the slope deviation between the orange and blue curves are likely due to the paraxial approximation made to compute the theoretical velocities. The quadratic term of STC should, therefore, be adjusted to get the requested velocity. Note that the end of the focal line also disturbs the measurement of the group velocity, which results in its sudden increase (purple curve in Fig.~\ref{fig:vel_int}). 

\section{Axiparabola with a constant energy focal line}

\begin{figure}[htbp]
    \centering
    \includegraphics[width=0.45\textwidth]{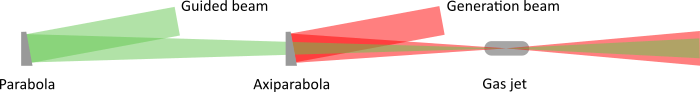}
    \caption{Example of setup for a guiding experiment. The generation beam is focused by an axiparabola and shot a few nanoseconds before the guided beam to allow the formation of the waveguide~\cite{smartsev2019axiparabola}.}
    \label{fig:guiding_exp}
\end{figure}

The great versatility of axiparabolas for applications provides the possibility to achieve various focal line distributions. Axiparabolas with various intensity distributions are of particular interest for the investigation of plasma channels generation for guiding purposes (Fig.~\ref{fig:guiding_exp}). The study of an axiparabola with another sag function also allows us to assess the validity and solidity of our theoretical model.  

Let's take the example of an axiparabola with a constant energy focal line. In order to obtain a focal line with a constant energy encircled in the central spot, the linear density of rays $\lambda_z$ needs to compensate for the first-zero radius decrease, as illustrated in Fig.~\ref{fig:int_int}. Therefore, following Eq. \eqref{eq:r_axi}, $\lambda_z$ needs to be proportional to the square of the incident rays radius on the axiparabola r : $\lambda_z\propto\left(r/R\right)^2$, which leads for a holed axiparabola to
\begin{align}
    f=f_0+\frac{1}{a}\,\ln\left(\frac{r}{R}e^{a\delta}\right)
\end{align}
with $a=\frac{1}{\delta}\,ln\left(\frac{R}{r_{hole}}\right)$ where $r_{hole}$ is the radius of the hole at the center of the axiparabola.

\begin{figure}[htbp]
    \centering
    \includegraphics[width=0.45\textwidth]{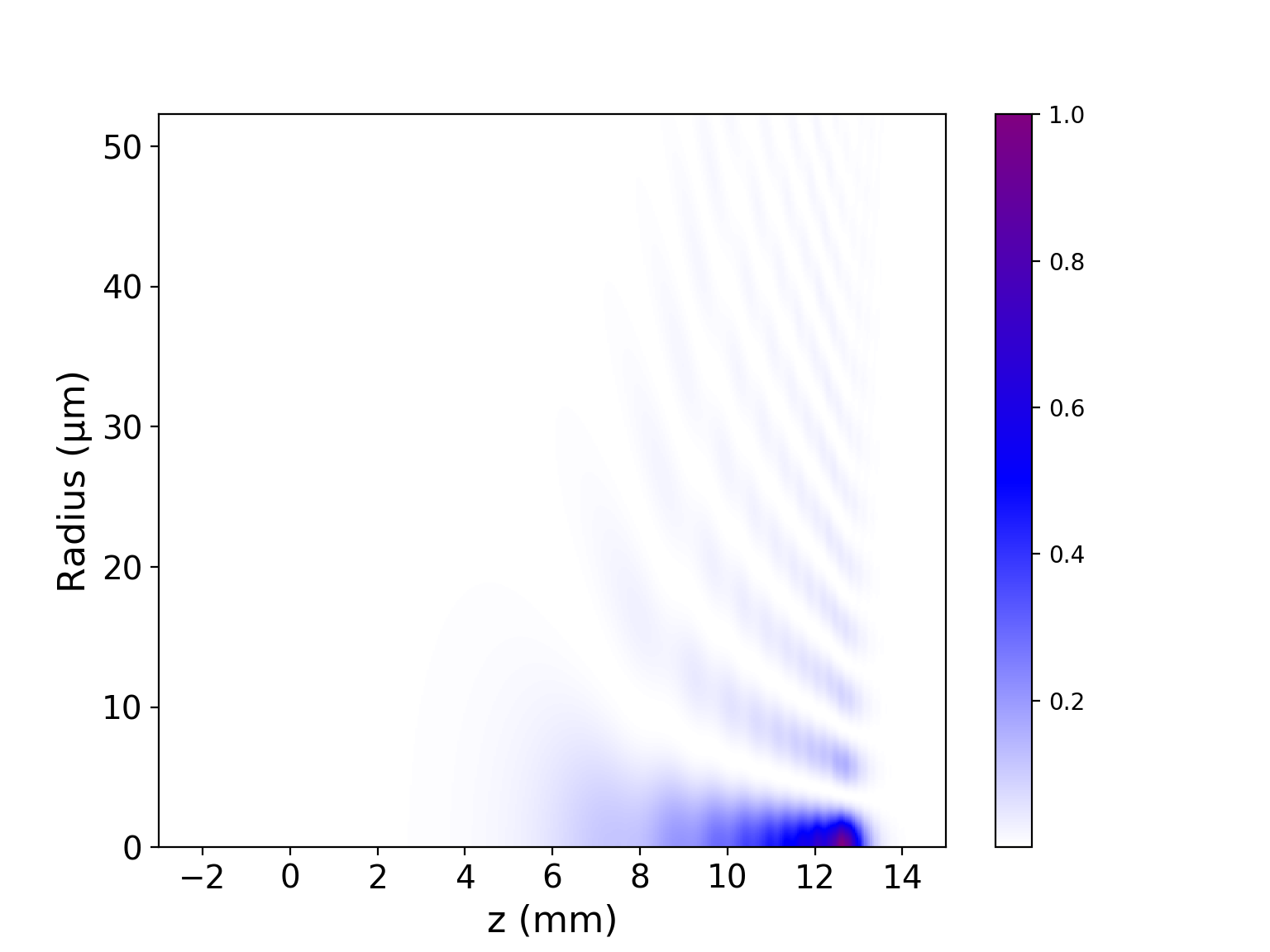}
    \caption{Intensity map of a constant energy focal line in arbitrary units as a function of $(r,z)$.}
    \label{fig:foc_en}
\end{figure}
\begin{figure}[h]
    \centering
    \includegraphics[width=0.45\textwidth]{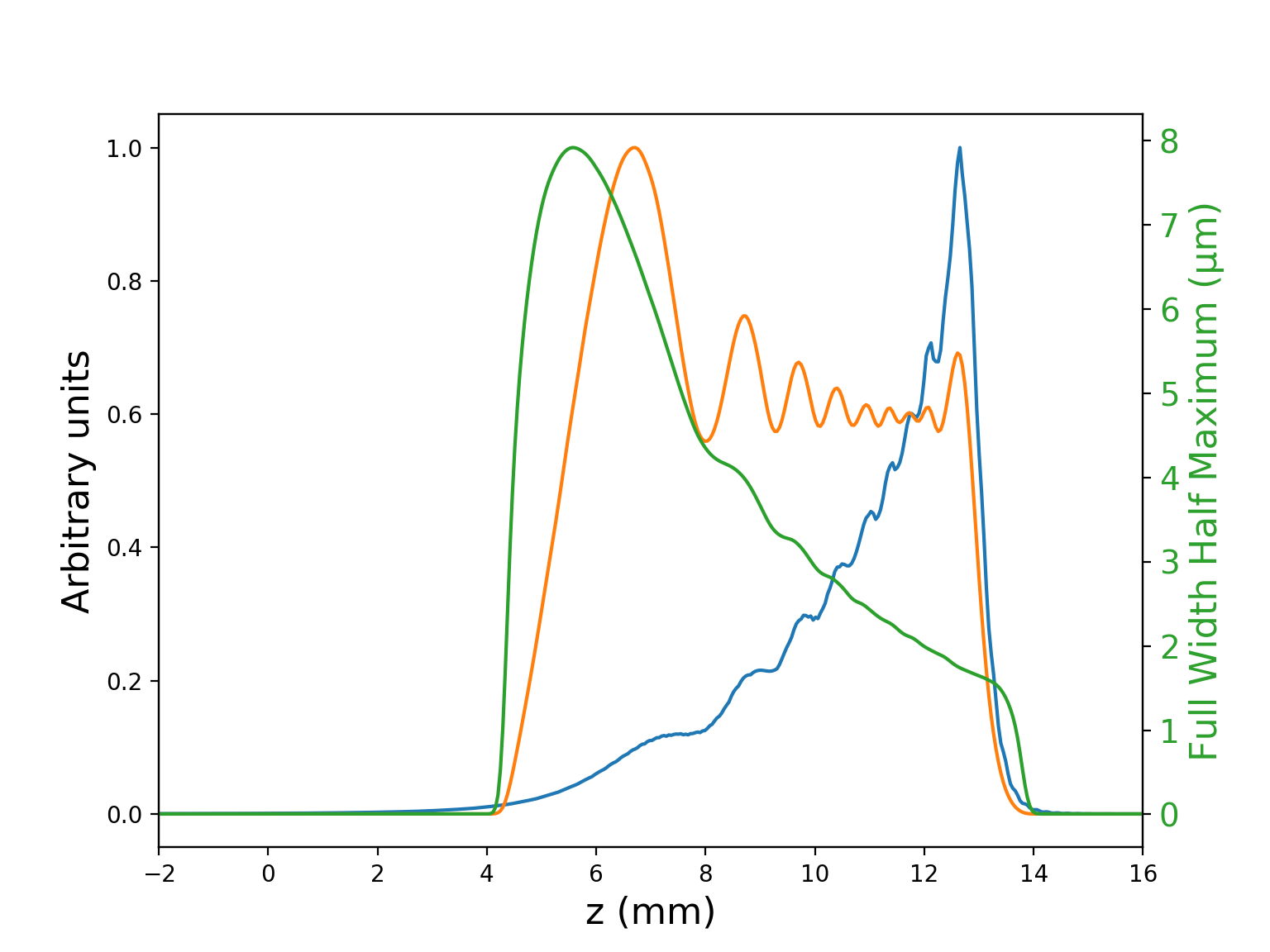}
    \caption{Variation along the focal line of the intensity (blue curve), energy (orange curve) and first-zero radius (green curve).}
    \label{fig:int_en}
\end{figure}
However, this simple model overestimates the energy for small radii, which would result in an increasing energy focal line. Therefore, we use an empirical formula similar to the previous one but more in adequation with reality for smaller radii.
In practice this condition is fulfilled for 
\begin{align}
    f=f_0+0.1\delta\frac{r}{R}+0.9\delta\left(\frac{r}{R}\right)^{\frac{1}{2}} \,,
\end{align}
leading to an expression for the group velocity:
\begin{align}
\frac{v}{c}=1+\frac{\delta^2}{f^2}\left(0.405\frac{r}{R}+0.135\left(\frac{r}{R}\right)^{\frac{3}{2}}+0.01\left(\frac{r}{R}\right)^2\right)\,.
\label{velocity_ener}
\end{align}

For simulations, we assume the same parameters as in Sec.~\ref{sec:cons_i}: a nominal focus $f_0=400$ mm, a focal depth $\delta=15$ mm and a radius $R=38.1$ mm.

\begin{figure}[htbp]
    \centering
    \includegraphics[width=0.45\textwidth]{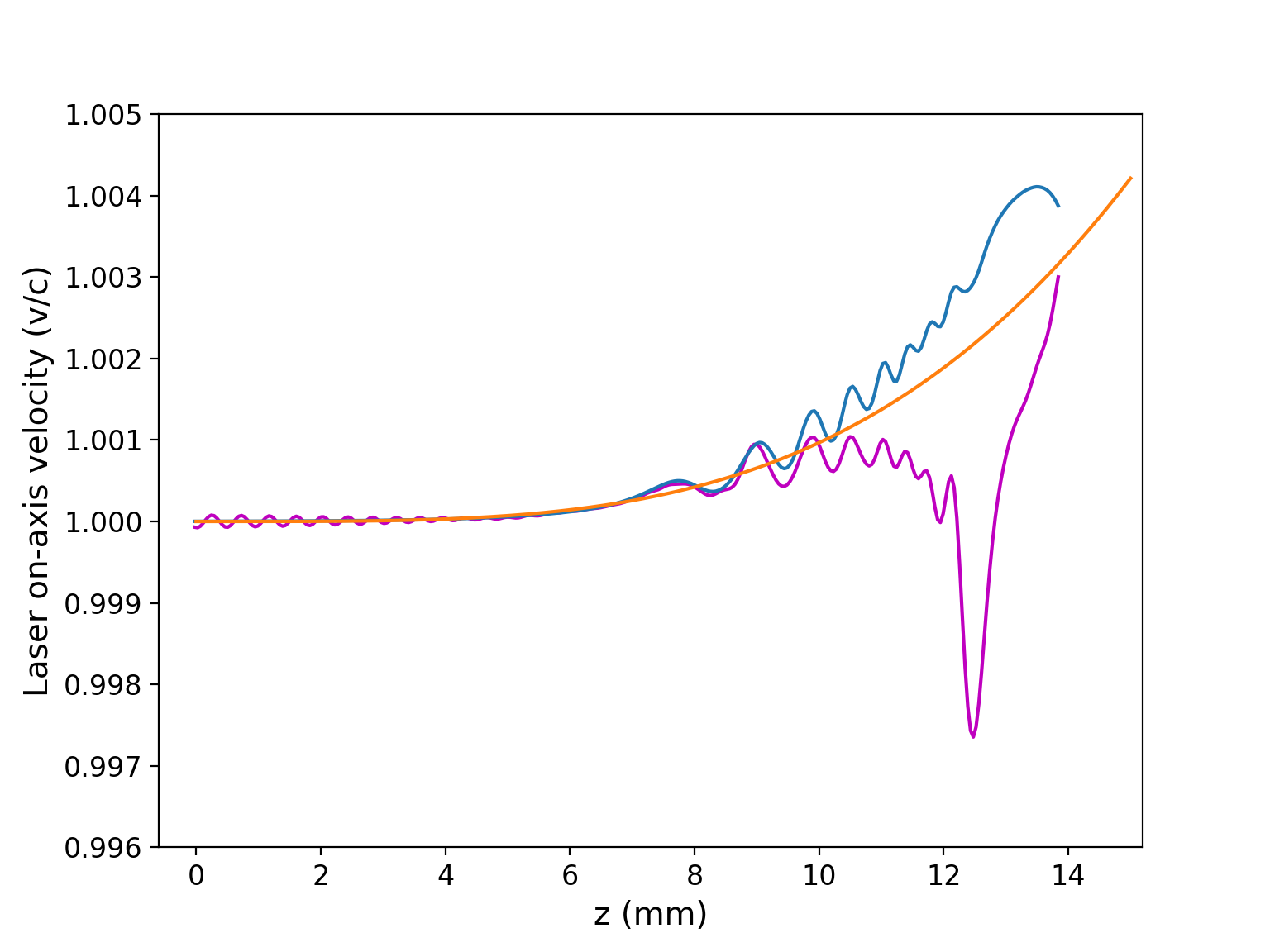}
    \caption{Group velocity as a function of the position along the focal line of the axiparabola. The orange curve results from  Eq.~\eqref{velocity_int}. The blue and purple curves correspond to simulation data  obtained  without and with the radial delay displayed in Fig~\ref{fig:ener_ret}, respectively.}
    \label{fig:vel_en}
\end{figure}
\begin{figure}[h]
    \centering
    \includegraphics[height=6cm]{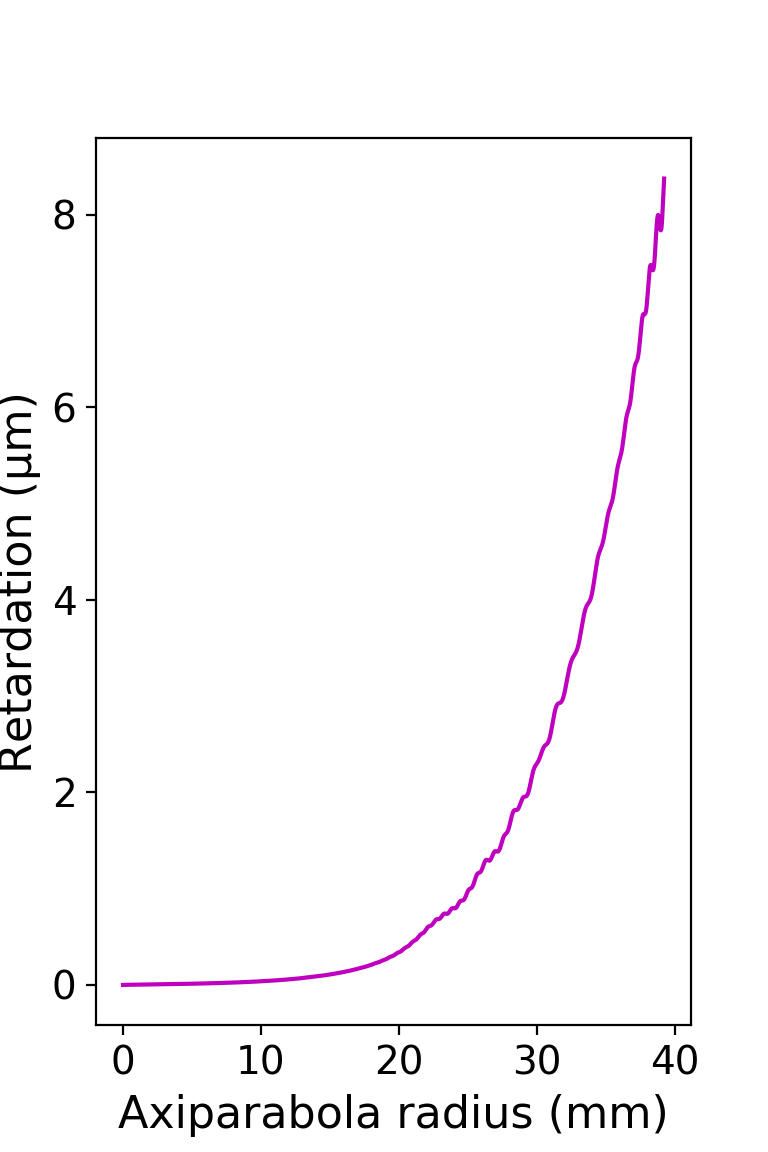}
    \caption{Pulse front delay required for the group velocity to  equal c, according to Eq.~\eqref{retardation_1}}
    \label{fig:ener_ret}
\end{figure}

We observe in  Fig.~\ref{fig:int_en} that the intensity effectively increases along the focal line to compensate for the decrease of the first-zero radius.

The theoretical group velocity matches the numerical estimate, with the deviation at the end of the focal line, which is mainly due to the ray approximation assumption made to derive equations. This  confirms the reliability of the simple model exposed in the first three sections, for axiparabolas with different sag functions and different purposes. The spatio-temporal couplings needed to obtain a focal line group velocity equal to c is also validated by simulation data. 

\section{Conclusion and future work}

In conclusion, we demonstrated the possibility to control the longitudinal intensity distribution and the beam velocity, over a distance much larger than the Rayleigh length, using an axiparabola. The adaptability of this aspheric mirror was illustrated by designing and presenting two optical configurations for different applications. We also showed through theory and simulations that the group velocity of the focal line can be controlled through spatio-temporal couplings  and that the corresponding  delay can be evaluated from the main axiparabola features. The  unique capabilities and versatility of axiparabolas open up new perspectives for manipulating intense and ultra-short laser pulse, which is a promising boost for the development of compact and flexible bright radiation and particles sources in laser wakefield acceleration frame. Moreover, a better control of these high intensity focal line properties (intensity distribution, propagation velocity) can also be an advantage for many other applications, e.g. soft X-ray laser~\cite{depresseux2015table}, pulse compression in a plasma~\cite{faure2005observation} or photon acceleration~\cite{wilks1989photon}.

\bibliography{bibliography}

\end{document}